%% file: main.tex
\documentclass[sigconf]{acmart}

\usepackage{natbib}
\input{tex/preamble}
\input{tex/figures}

\input{tex/tables}
\title[Alpha-GPT v2]{Alpha-GPT 2.0: Human-in-the-Loop AI for Quantitative Investment}

\begin{document}
\bibliographystyle{unsrt}

\newcommand{\HKUST}{
\institution{The Hong Kong University of Science and Technology}
\country{Hong Kong SAR}
}

\newcommand{\HKUSTGZ}{
\institution{The Hong Kong University of Science and Technology (Guangzhou)}
\country{China}
}

\newcommand{\Columbia}{
\institution{Columbia University}
\country{USA}
}

\newcommand{\IDEA}{
\institution{IDEA Research}
\country{China}
}

\author{Hang Yuan}
\authornote{Work done during internship at IDEA Research.}
\affiliation{\HKUST}
\email{hyuanak@connect.ust.hk}

\author{Saizhuo Wang}
\authornotemark[1]
\affiliation{\HKUST}
\email{swangeh@connect.ust.hk}


\author{Jian Guo}
\authornote{Corresponding author}
\affiliation{\IDEA}
\email{guojian@idea.edu.cn}

\input{tex/abstract}
\maketitle
\pagestyle{plain}

\input{tex/intro}
\input{tex/workflow}

\bibliography{references}

\end{document}

%% file: tex/preamble.tex
\usepackage{graphicx} 
\usepackage{hyperref}
\usepackage[linesnumbered,ruled,vlined]{algorithm2e}
\usepackage{enumitem}
\usepackage{booktabs}
\usepackage{array}
\usepackage{graphicx}   
\usepackage{subcaption} 
\usepackage{tabularx}
\usepackage{xcolor}

\setcopyright{none}
\renewcommand\footnotetextcopyrightpermission[1]{} 

%% file: tex/figures.tex
\newcommand{\figureAlphaGPTIntro}{
\begin{figure*}[!ht]
    \centering
    \includegraphics[width=\textwidth]{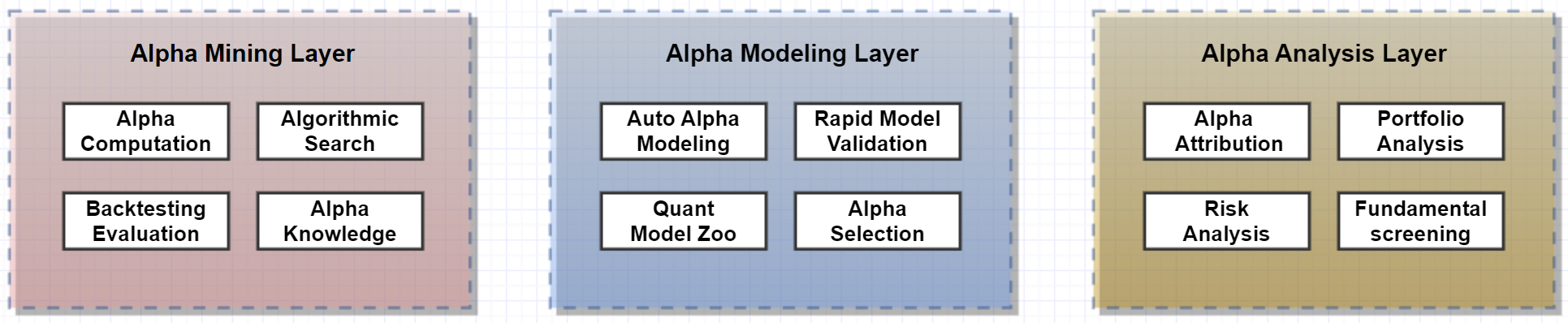}
    \caption{Overview of Alpha-GPT 2.0.}
    \label{fig:intro}
\end{figure*}
}

\newcommand{\figureAlphaGPTFrame}{
\begin{figure*}[!t]
    \centering
    \includegraphics[width=0.9\textwidth]{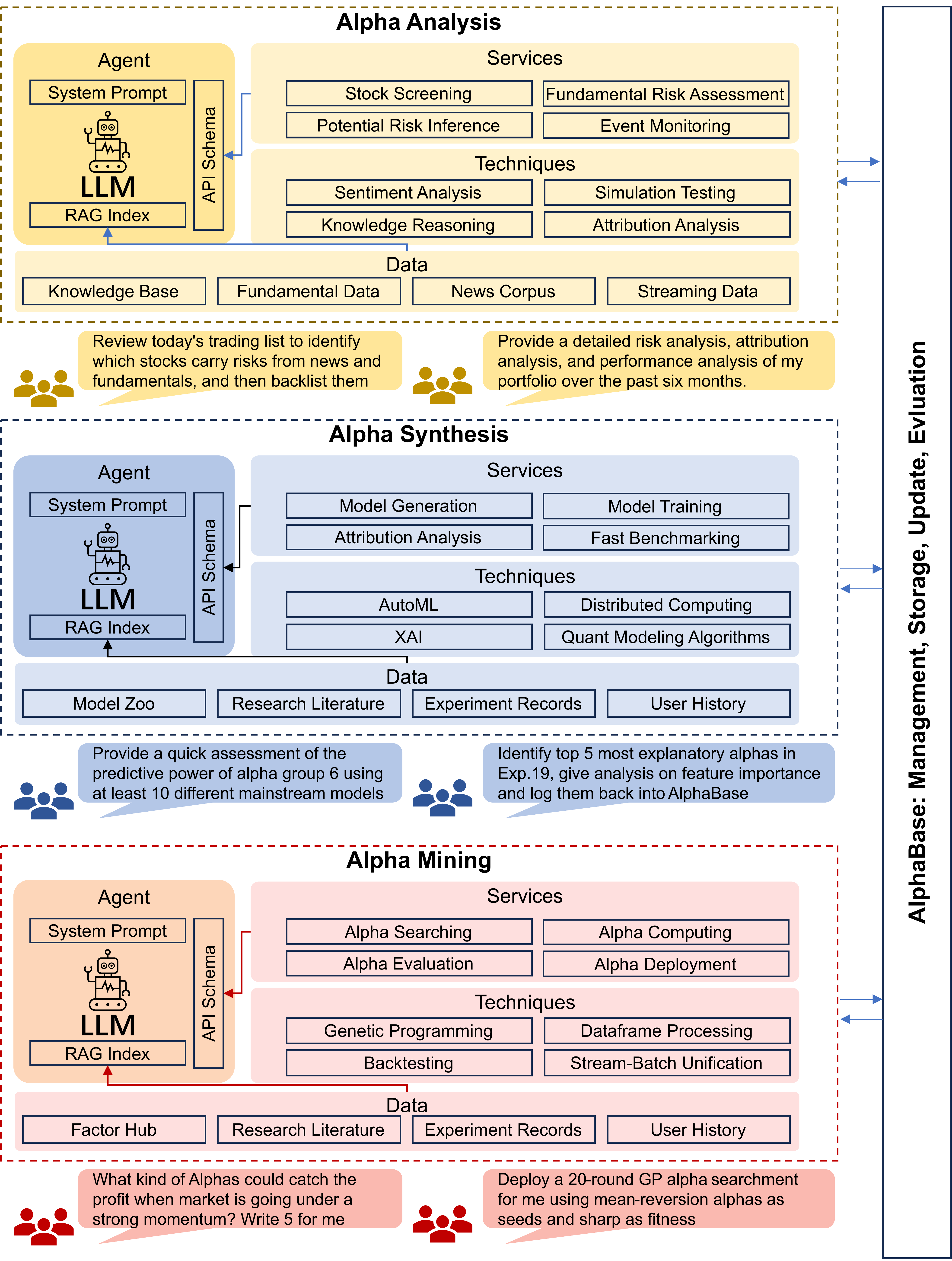}
    \caption{Framework of Alpha-GPT 2.0}
    \label{fig:frame}
\end{figure*}
}

%% file: tex/abstract.tex
\begin{abstract}
Recently, we introduced a new paradigm for alpha mining in the realm of quantitative investment, developing a new interactive alpha mining system framework, Alpha-GPT. This system is centered on iterative Human-AI interaction based on large language models, introducing a Human-in-the-Loop approach to alpha discovery. In this paper, we present the next-generation Alpha-GPT 2.0 \footnote{Draft. Work in progress}, a quantitative investment framework that further encompasses crucial modeling and analysis phases in quantitative investment.
This framework emphasizes the iterative, interactive research between humans and AI, embodying a Human-in-the-Loop strategy throughout the entire quantitative investment pipeline.
By assimilating the insights of human researchers into the systematic alpha research process, we effectively leverage the Human-in-the-Loop approach, enhancing the efficiency and precision of quantitative investment research.
\end{abstract}

%% file: tex/intro.tex
\section{Introduction}
A standard quantitative investment pipeline commences with the mining of alphas \cite{tulchinsky_introduction_2019, kakushadze_101_2016} that encapsulate valuable market information. These alphas function as features in predictive models, forecasting future stock returns over a specific period. Following the creation of these models, asset portfolio optimization \cite{lai_survey_2022} occurs, formulating trading positions and strategies based on the synthesized alpha. Subsequently, the developed strategy undergoes comprehensive analysis and rigorous testing before evaluation and consequent deployment in live trading. This systematic approach, responsive to new data and market conditions, exemplifies a refined quantitative investment strategy.

\figureAlphaGPTIntro

Traditionally, the quantitative investment workflow is a collaborative process that requires specialized researchers at each stage. However, due to a lack of experienced and skilled researchers, scaling up the output of quantitative research can be a formidable challenge. With the advent of technologies like automated alpha discovery algorithms \cite{yu_generating_2023, zhang_autoalpha_2020} and AutoML \cite{he_automl_2021}, many institutions have begun adopting AI-based automated processes for quantitative investment research. While this progression marks a significant milestone, it also presents notable problems. Primarily, these AI algorithms are compute-intensive. As research advances, computational power tends to yield diminishing returns in terms of strategy performance improvement, making the approach increasingly cost-inefficient.

To address these challenges, we propose a new paradigm for the quantitative investment research workflow, namely Human-In-The-Loop AI for Quantitative Investment. This innovative approach combines the market insights, understanding, and experience of human researchers with the efficiencies of an AI-based automated quantitative investment research system. Through iterative rounds of collaboration across crucial stages of the research process, this dynamic interactive approach facilitates the effective discovery of trading alphas and investment strategies. It leverages the insights and experience of human researchers to guide the systems such as in alpha mining and model searching process more effectively. Simultaneously, it uses the experimental results of the automated research system to continuously inspire human researchers to better understand financial markets. This cyclical, multi-round interactive process allows for efficient and significantly effective exploration in quantitative investment research. The proposed paradigm underscores the potential of a synergistic human-AI collaboration in advancing the field of AI-based quantitative investment.

To implement this Human-in-the-Loop paradigm for quantitative investment research, we have adopted a Multi-agent architecture, termed \textbf{Alpha-GPT 2.0}. This architecture employs specialized agents, each trained and developed to excel in distinct segments of the quantitative investment workflow, such as alpha mining, alpha modeling, and alpha analysis. These agents are powered by large language models and serve a dual purpose: they efficiently drive and interact with the various modules and tools within the existing automated quantitative investment algorithms, and they accurately interpret the research intentions and operational instructions of human researchers. The agents, each focused on a specific stage of the workflow, connect the whole quantitative investment process into a research cycle. This cycle includes mining alphas from data, combining alphas into models, conducting comprehensive analysis of research results, and based on the understanding and analysis of these results, informing the direction for the next round of alpha researching. In a innovative approach, these AI agents—each tailored for distinct stages of the quantitative investment process—form a collaborative multi-agent system. They synergize to complete the research cycle. Importantly, at every stage, human researchers can infuse their insights and ideas into the research cycle through their interactions with the respective agents. Concurrently, the system's feedback on the experiments allows human researchers to conveniently and comprehensively receive an extensive understanding of the experiments results, thereby guiding the direction of subsequent experiments.
\figureAlphaGPTFrame

Our contributions in this work can be summarized from these standpoints:
\begin{itemize}[leftmargin=*]
    \item We extended the paradigm proposed in Alpha-GPT \cite{wang_alpha-gpt_2023} to full-pipeline automation, which synergizes human expertise and AI capabilities to enhance the efficiency and effectiveness of quantitative investment research.
    \item We established a comprehensive research workflow in Alpha-GPT 2.0, integrating alpha mining, modeling and analysis with iterative feedback, enriching quantitative investment research.
\end{itemize}


%% file: tex/workflow.tex
\section{System Workflow and Architecture}
The workflow of Alpha-GPT 2.0 encompasses the three most crucial phases of quantitative investment research: Alpha Mining, Alpha Modeling, and Alpha Analysis. In each layer, we constructed an LLM (Large Language Model) powered autonomous agent, tailored to the specific tasks of that layer. These agents are equipped with a variety of tools necessary for accomplishing their respective tasks, such as calling APIs to deploy an algorithmic alpha search experiment or training a batch of ML models to benchmark the predictive ability of selected alpha factors. Furthermore, each layer's agents are equipped with a collection of memories necessary for completing their investment research tasks, for instance, a detailed annotated alpha base or users' past chat history and use logs for LLM reflection. Typically, the workflow of alpha investment research is predefined, and we do not require LLM for complex tasks' planning. We have prepared standard SOPs (Standard Operating Procedures) for each layer of the investment research workflow to help LLM clearly understand the trajectory of their tasks. Common task flows like assisting users in extracting alpha factors from diverse market data, combining these factors through machine learning models to create a synthesized alpha signal also providing interpretations of these factors, constructing a trading list by portfolio optimization using the synthesized alpha, and giving comprehensive analysis and stock filtering based on fundamental and news event data. The subsequent sections will provide detailed descriptions of each of these three critical components.

\subsection{Alpha Mining Layer}
In the Alpha-GPT 1.0 paper, the design of the Alpha Mining Layer was elaborated in detail, particularly focusing on the significant challenge of using LLMs for interpreting trading ideas. The primary objective of this layer is to assist users in more efficiently mining alpha factors from diverse market data. Specifically, its key tasks include helping users translate market insights and trading ideas from natural language into expression-based alpha factors, and enhancing or exploring these alpha factors by deploying algorithmic alpha search experiments. 
The alpha mining agent at this layer, equipped with a comprehensive set of tools, facilitates tasks ranging from alpha computation, back-testing evaluation, algorithmic search enhancement, alpha deployment to maintaining the alpha base. In fact, due to these comprehensive tools for various types of tasks, the agent also assists users as copilots in various subtasks, such as conducting a back-test on alpha factors and generating analysis reports, or extracting alpha factors with specific tags from the alpha base to deploy a GP algorithm search. To enable the agent to effectively complete different alpha mining tasks, various types of data and knowledge have been prepared as the agent's memory. These include an alpha base with detailed annotations on the style, market characteristics, and user descriptions of rich alpha factors; a wealth of processed literature on financial signal research, such as books, research reports, and papers; as well as records and logs of experiments, users' chat histories, and comments from the research process.

\subsection{Alpha Modeling Layer}
The Alpha Modeling Layer is primarily focuses on ML/DL modeling of alpha factors to construct effective predictive alpha signals, optimize investment portfolios, and assist infeature selection and interpretation. This layer provides the agent with a  full suite of tool interfaces necessary for model construction, including model training / testing, model structure searching, hyperparameter optimization, model benchmarking and comparision, feature selection and explanation, as well as subsequent investment portfolio optimization and trade list generation. The agent utilizes these tools by selecting various actions to support the external functionalities required in the modeling phase. A rich model zoo is available in the memory, containing knowledge on model characteristics and templates for model parameter configuration. After retrieving the relevant model information, the LLM can effectively generate the configs for model construction and training, enabling subsequent model experiment deployment. Moreover, the accumulation of experimental log information and the user's dialogue history during the modeling process can also aid the agent in optimizing its work in this layer
In practical scenario, it's often necessary to conduct a rapid assessment of the predictive capability of a combination of alpha factors, providing a basic benchmark of predictive effectiveness. For instance, when a user asks, 'Provide me a quick assessment of the predictive capability for week's return of alpha factors group 6 using at least 10 mainstream ML models,' the agent executes a simple planning process. It begins by retrieving the specified batch of alpha factors from the alpha base, then prepares the necessary data for the experiment, including the labels. Utilizing the knowledge of machine learning models and basic model hyperparameter templates present in the prompt context, the LLM generates the required configurations for training ten mainstream machine learning models. Subsequently, it builds the models in batch by calling the APIs for deploying model training experiments in this layer, followed by evaluating and testing the training results. Finally, it provides a report detailing the benchmark of predictive capability and relays this feedback to the user through the agent.

\subsection{Alpha Analysis Layer}
In previous layers, the focus was on mining alpha factors, training models, and building investment portfolios  based on price-volume type market data. However, in real-world investing, it's crucial to incorporate real-time information on companies, industries, and market fundamentals, as well as news events, to make robust investment recommendations and avoid risks to enhance portfolio performance. The alpha analysis agent is designed to provide multi-faceted portfolio analysis. This includes blacklisting stocks with fundamental and event-driven risks for investment filtering and delivering in-depth market analysis reports on fundamentals, effectively identifying market investment logic, current industry and capital flow trends, and potential financial risks. A key feature of Alpha Analysis Layer is the construction of a large-scale financial behavior knowledge graph, which forms an integral part of the agent's memory. It links billions of financial entities and events from extensive financial big data, including sectors like industry, supply, capital, and litigation chains, facilitating efficient market logic mining and risk prediction. This layer's agent employs Think-on-Graph inference technology, leveraging the vast financial knowledge graph and structured financial texts to drive the LLM's deep reasoning. This approach ensures that the financial analysis tasks are reliable, controllable, interpretable, and traceable.
For instance, when users generate an investment portfolio based on alpha signals, there are inherent risks from market fundamentals and events that need to be considered. Users can instruct the agent to 'Review today's trading list and filter out stocks carrying risks and blacklist them.' The agent first accesses today's optimized portfolio list, then extracts relevant knowledge of these companies from the memory,  including data from the financial behavior knowledge graph, followed by risk assessment and analysis through the LLM. It would identify and exclude stocks with significant risks, finally blacklisting them and reporting back to the user. Similarly, in industry analysis scenarios, users may ask the agent to 'Analyze the current state and risks of all primary industries and reduce the weight of stocks in my investment holdings that have significant industry risks to half.' The agent then extracts industry knowledge from relevant databases and industry chain graphs, provides it to the LLM for analysis and risk assessment, delivers a comprehensive report to the user, and adjusts the portfolio accordingly by halving the weight of stocks in high-risk industries.